\documentstyle[multicol,aps,psfig]{revtex}

\begin{document}
\preprint{LBNL-60015}

\title{Interference effect in elastic parton energy loss in a finite medium}
\author{Xin-Nian Wang}
\address{Nuclear Science Division, MS 70-319,
Lawrence Berkeley National Laboratory, Berkeley, CA 94720 USA}

\date{\today}

\maketitle

\vspace{-1.5in}
{\hfill LBNL-58446}
\vspace{1.4in}

\begin{abstract}
Similar to the radiative parton energy loss due to gluon bremsstrahlung, 
elastic energy loss of a parton undergoing multiple scattering in a 
finite medium is demonstrated to be sensitive to interference effect.
The interference between amplitudes of elastic scattering via a gluon
exchange and that of gluon radiation reduces the effective elastic
energy loss in a finite medium and gives rise to a non-trivial length
dependence. The reduction is most significant for a propagation
length $L< 4/\pi T$ in a medium with a temperature $T$. Though the 
finite size effect is not significant for the average parton propagation
in the most central heavy-ion collisions, it will affect the centrality
dependence of its effect on jet quenching.

\end{abstract}

\begin{multicols}{2}


One of the remarkable phenomena observed in central nucleus-nucleus collisions
at the Relativistic Heavy-ion Collider (RHIC) is jet quenching as manifested
in the suppression of high transverse momentum hadron 
spectra \cite{phenix-r1,star-r1}, azimuthal angle 
anisotropy \cite{star-jetv2} and suppression of the away-side
two-hadron correlations \cite{star-jet}. The observed patterns 
of jet quenching, their centrality, momentum and colliding energy 
dependence in light hadron spectra and correlations are consistent 
with the picture of parton energy loss \cite{Wang:2003mm}. 
However, recent experimental data on measurements of single
non-photonic electron spectra \cite{Adler:2005xv}
seem to indicate a suppression of heavy quarks that is not consistent
with the theoretical predictions based on current implementation of
heavy quark energy 
loss \cite{Djordjevic:2003qk,Armesto:2005mz,vanHees:2005wb}. 
This has led to a renewed interest in the
elastic (or collisional) energy loss and its effects in the observed
jet quenching phenomena \cite{Mazumder,Mustafa:2004dr,Wicks:2005gt}.

Elastic energy loss by a fast propagating parton in general is caused by
its elastic scattering with thermal partons in the medium through 
one-gluon exchange. The gluon exchange can be considered as the emission
and subsequent absorption of gluons by two scattering partons and therefore
should also be subject to formation time of the virtual gluon due to 
interference. In a finite medium when the propagation length
is comparable to the formation time, the interference should reduce
the effective elastic energy loss as compared to an infinite medium.
Since the typical energy exchange with a thermal parton is $\omega\sim T$,
the average formation time is therefore controlled by the temperature of
the medium. In a recent study by Peigne {\it et al}. \cite{Peigne:2005rk}, 
however, collisional energy loss was found to be suppressed significantly 
for considerably large medium size. This might be partially due to
the complication of subtraction of induced radiation associated 
with the acceleration of color charges within a finite period of time 
in the semi-classical approach.

For partons produced via a hard process that go through
further multiple scattering, elastic scattering via one-gluon
exchange often bears many similarities to the radiative (or gluon
bremsstrahlung) processes. In this paper, we will consider the elastic 
scattering within the same framework of multiple parton scattering 
that was employed to study the radiative energy 
loss \cite{wgdis,Zhang:2003yn}. We will show that
it is the same interference between the amplitudes of 
elastic scattering and induced radiation that leads to the
reduction of the elastic energy loss. Such a reduction gives
rise to a non-trivial medium size dependence of the elastic energy loss.

For the purpose of illustration and derivation, we start with the
double quark scattering processes in deeply inelastic scattering (DIS)
off a nucleus and their effects on the nuclear modification of the
effective quark fragmentation functions. 
The differential cross section of semi-inclusive processes 
$e(L_1) + A(p) \longrightarrow e(L_2) + h (\ell_h) +X$ in DIS can be
expressed in general as
\begin{equation}
E_{L_2}E_{\ell_h}\frac{d\sigma_{\rm DIS}^h}{d^3L_2d^3\ell_h}
=\frac{\alpha^2_{\rm EM}}{2\pi s}\frac{1}{Q^4} L_{\mu\nu}
E_{\ell_h}\frac{dW^{\mu\nu}}{d^3\ell_h} \; ,
\label{sigma-dis}
\end{equation}
in terms of the semi-inclusive hadronic tensor,
\begin{eqnarray}
E_{\ell_h}\frac{dW^{\mu\nu}}{d^3\ell_h}&=&
\frac{1}{2}\sum_X \langle A|J^\mu(0)|X,h\rangle 
\langle X,h| J^\nu(0)|A\rangle \nonumber \\
&\times& 2\pi \delta^4(q+p-p_X-\ell_h) \;,
\end{eqnarray}
and the leptonic tensor,
$L_{\mu\nu}=\frac{1}{2}\, {\rm Tr}(\gamma \cdot L_1 \gamma_{\mu}
\gamma \cdot L_2 \gamma_{\nu})$ ,
where $q=[-Q^2/2q^-, q^-, \vec{0}_{\perp}]$ is the four-momentum of the
virtual photon, $p=[p^+, 0, \vec{0}_{\perp}]$ is the momentum of the 
target per nucleon, $s=(p+L_1)^2$ is the total invariant mass of the 
lepton-nucleon system, $J_\mu$ is the hadronic electromagnetic (EM) 
current, $J_\mu=e_q \bar{\psi}_q \gamma_\mu\psi_q$, and $\sum_X$ runs 
over all possible intermediate states.

The leading-twist contribution to the semi-inclusive hadronic
tensor to the lowest order in the strong coupling constant comes
from a single virtual photon and quark scattering,
\begin{equation}
\frac{dW_{\mu\nu}^{S(0)}}{dz_h}
= \sum_q f_q^A(x) H^{(0)}_{\mu\nu}(x,p,q) D_{q\rightarrow h}(z_h)\ ,
\label{eq:w-s}
\end{equation}
where $f_q^A(x)$ is the quark distribution in the nucleus,
$x=Q^2/2p^+q^-$ the Bjoken variable, $D_{q\rightarrow h}(z_h)$ the 
quark fragmentation function and
\begin{equation}
H^{(0)}_{\mu\nu}(x,p,q)=e_q^2\frac{\pi}{2p^+q^-}\, 
{\rm Tr}(\gamma \cdot p \gamma_{\mu} \gamma \cdot(q+xp) \gamma_{\nu})
\end{equation}
the hard part of $\gamma^* +q$ partonic scattering.

\begin{figure}
\centerline{\psfig{figure=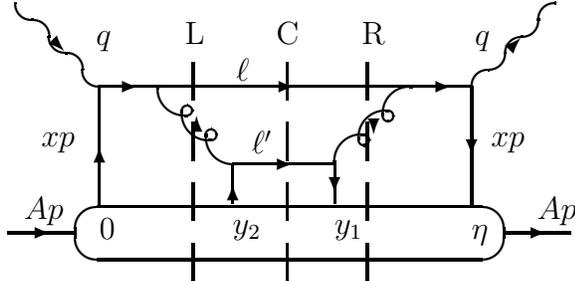,width=3.0in,height=1.5in}}
\caption{Cut diagrams for double quark-quark scattering in DIS off a nucleus.}
\label{fig1}
\end{figure}

For the simplest case, let us consider multiple scattering between two quarks
with different flavors in DIS off a large nucleus as illustrated in
Fig.~\ref{fig1} (with the central cut). In this process, a quark ($q$) knocked 
out by the virtual photon undergoes a secondary scattering with another 
quark $q^\prime$ from the nucleus. It contributes to the semi-inclusive 
DIS at twist four similarly
as multiple scattering with gluons in the discussion of induced gluon 
radiation. We assume the flavors of the two quarks to be different so 
that there is no contribution from crossing diagrams. 
The contribution from the central-cut diagram in Fig.~\ref{fig1} to 
the semi-inclusive tensor is, as given in Ref.~\cite{wgdis},
\begin{eqnarray}
\frac{dW^{D(C)}_{\mu\nu}}{dz_h}&=& 
\frac{\alpha_s C_F }{2\pi}\int\frac{d\ell_T^2}{\ell_T^2} 
\int_{z_h}^1\frac{dz}{z} \left[ D_{q\rightarrow h}(z_h/z)
\frac{1+z^2}{(1-z)^2} \right. \nonumber \\
&+& \left. D_{q^\prime \rightarrow h}(z_h/z) \frac{1+(1-z)^2}{z^2} \right]
\nonumber \\
& \times &\frac{x}{Q^2}
\frac{2\pi\alpha_s}{N_C} T^{A(C)}_{qq^\prime}(x,x_L)
H^{(0)}_{\mu\nu}(x,p,q)\, ,
\end{eqnarray}
where
\end{multicols}
\begin{eqnarray}
T^{A(C)}_{qq^\prime}(x,x_L)&=&
p^+\int\frac{d\eta^-}{2\pi}dy_1^-dy_2^- e^{i(x+x_L)p^+\eta^-} 
(1-e^{-ix_Lp^+y_2^-})
(1-e^{-ix_Lp^+(\eta^--y_1^-)}) 
\theta(-y_2^-)\theta(\eta^--y_1^-) \nonumber \\
&\times&\langle A|\bar{\psi}_q(0)\frac{\gamma^+}{2}\psi_q(\eta^-) 
\bar{\psi}_{q^\prime}(y_1^-)\frac{\gamma^+}{2}\psi_{q^\prime}(y_2^-)|A\rangle
\, .
\end{eqnarray}
\begin{multicols}{2}
\noindent
is the two-quark correlation function in the nucleus. The four terms with 
different phase factors in the above matrix element correspond to 
hard-soft, double hard quark scattering and their interferences,
similarly to the Landau-Pomeranchuck-Migdal (LPM) interference in
the processes of induced gluon radiation \cite{wgdis}.

In the hard-soft quark
scattering, the first or leading quark that was knocked out of the 
nucleus by the hard virtual photon becomes off-shell. It then 
radiates a real gluon which interacts with another soft quark that
carries zero fractional momentum ( + component) in the collinear 
limit (neglecting the initial transverse momentum of the quark)
and converts it into the final quark with momentum $\ell^\prime$.
In this 
case, the final total longitudinal fractional momentum of the two quarks
$x_L=(\ell^+ + \ell^{\prime+})/p^+=\ell_T^2/2p^+q^-z(1-z)$ comes from
the initial leading quark before its interaction with the virtual photon.
This kinematics corresponds to the induced emission of a secondary quark
like the case of induced gluon radiation. Note that quarks from the
matrix elements normally are off-shell and carry transverse momentum.
The kinematics in the hard-soft quark scattering process will then
allow both finite fractional longitudinal and transverse momentum
when the secondary quark is off-shell. It is only in the collinear
approximation, {\em i.e.}, neglecting the initial transverse momentum,
that the quark lines from the nucleus can be considered on-shell.
The kinematics in the hard-soft processes then requires the secondary
quark to carry zero fractional longitudinal momentum.

In the process of double hard quark scattering, the leading quark becomes
on-shell after its interaction with the virtual photon. It then
scatters with another quark that carries finite momentum fraction $x_L$.
Therefore, the final total longitudinal 
fractional momentum of the two quarks $x_L$ is transferred from the
initial momentum of the second quark. This is equivalent to elastic
scattering between two quarks via one-gluon exchange.

For a complete calculation of double quark scattering processes in DIS,
one also has to include interference between triple quark scattering 
and single quark scattering as given by the left and right cut diagrams 
in Fig.~\ref{fig1}. Their contributions to the semi-inclusive
hadronic tensor are very similar to the central-cut diagram,
\begin{eqnarray}
\frac{dW^{D(R,L)}_{\mu\nu}}{dz_h}&=& 
\frac{\alpha_s C_F }{2\pi}\int\frac{d\ell_T^2}{\ell_T^2} 
\int_{z_h}^1\frac{dz}{z} \left[ D_{q\rightarrow h}(z_h/z)
\frac{1+z^2}{(1-z)^2} \right. \nonumber \\
&+& \left. D_{g \rightarrow h}(z_h/z) \frac{1+(1-z)^2}{z^2} \right]
\nonumber \\
& \times &\frac{x_B}{Q^2}
\frac{2\pi\alpha_s}{N_C} T^{A(R,L)}_{qq^\prime}(x,x_L)
H^{(0)}_{\mu\nu}(x,p,q)\, ,
\end{eqnarray}
with the corresponding four-quark matrix elements
\end{multicols}
\begin{eqnarray}
T^{A(R)}_{qq^\prime}(x,x_L)&=&
p^+\int\frac{d\eta^-}{2\pi}dy_1^-dy_2^- e^{i(x+x_L)p^+\eta^-}
(-)(1-e^{-ix_Lp^+y_2^-}) 
\theta(-y_2^-)\theta(y_2^--y_1^-) \nonumber \\
&\times&\langle A|\bar{\psi}_q(0)\frac{\gamma^+}{2}\psi_q(\eta^-) 
\bar{\psi}_{q^\prime}(y_1^-)\frac{\gamma^+}{2}\psi_{q^\prime}(y_2^-)|A\rangle
\, , \\
T^{A(L)}_{qq^\prime}(x,x_L)&=&
p^+\int\frac{d\eta^-}{2\pi}dy_1^-dy_2^- e^{i(x+x_L)p^+\eta^-} 
(-)(1-e^{-ix_Lp^+(\eta^--y_1^-)})
\theta(y_1^--y_2^-)\theta(y^--y_1^-) \nonumber \\
&\times&\langle A|\bar{\psi}_q(0)\frac{\gamma^+}{2}\psi_q(\eta^-) 
\bar{\psi}_{q^\prime}(y_1^-)\frac{\gamma^+}{2}\psi_{q^\prime}(y_2^-)|A\rangle
\, .
\end{eqnarray}
\begin{multicols}{2}
\noindent
Note the negative signs before the phase factors and the different time 
orderings as given by the $\theta$ functions in the matrix elements
from the interference contributions. In addition to the three cut
diagrams in Fig.~\ref{fig1}, one should also include virtual corrections
which correspond to diagrams with cut quark lines right after the virtual
photon interaction. The total contribution to the 
semi-inclusive spectra from double quark scattering should include 
all three cut diagrams and virtual corrections from which one can 
then calculate the nuclear modification to the effective 
fragmentation function due to double quark scattering. 
The virtual corrections make the total
contribution from double quark scattering to the modified fragmentation 
functions unitary. However, they do not contribute to the quark
energy loss we discuss in this paper.

To compute the energy loss experienced by the leading quark due to
double quark scattering, one can identify it with the energy fraction 
the leading quark loses to the second quark in the central-cut
diagram or the radiated gluon in the left and right cut diagrams.
When summing these contributions, one encounters the combination
of the three $\theta$-functions,
\end{multicols}
\begin{eqnarray}
\int dy_1^-dy_2^-\left[\theta(-y_2^-)\theta(y_2^--y_1^-)
+\theta(\eta^--y_1^-)\theta(y_1^--y_2^-)-\theta(-y_2^-)\theta(\eta^--y_1^-)
\right]
=\int_0^{\eta^-}dy_1^-\int_0^{y_1^-}dy_2^- \; , \label{eq:theta}
\end{eqnarray}
\begin{multicols}{2}
\noindent
which is a path-ordered integral limited by the value of $\eta^-\sim 1/xp^+$.
We will neglect any such contact contributions as compared to
the dominant contributions whose integration over the location
of the second quark is only limited by the size of the nucleus.
In this case the two quarks come from two different nucleons in 
the nucleus and the corresponding contributions are enhanced by
$A^{1/3}$ due to the large nuclear size as compared to the contact
contribution. The dominant quark energy loss due to double quark
scattering is then
\begin{eqnarray}
\Delta z_{qq} =\frac{\alpha_s^2C_F}{2N_c}
\int\frac{d\ell_T^2}{\ell_T^2} \int dz \frac{1+(1-z)^2}{z p^+q^-} 
\frac{T^{A}_{qq^\prime}(x,x_L)}{f_q^A(x)},
\label{eq-eloss1}
\end{eqnarray}
where the four-quark matrix element $T^{A}_{qq^\prime}(x,x_L)$
is defined as
\end{multicols}
\begin{eqnarray}
T^{A}_{qq^\prime}(x,x_L)&=&
p^+\int\frac{d\eta^-}{2\pi}dy_1^-dy_2^- e^{ixp^+\eta^- +ix_Lp^+(y_2^- -y_1^-)} 
(1-e^{-ix_Lp^+y_2^-})
\theta(-y_2^-)\theta(\eta^--y_1^-) \nonumber \\
&\times&\langle A|\bar{\psi}_q(0)\frac{\gamma^+}{2}\psi_q(\eta^-) 
\bar{\psi}_{q^\prime}(y_1^-)\frac{\gamma^+}{2}\psi_{q^\prime}(y_2^-)|A\rangle
\, .
\end{eqnarray}
\begin{multicols}{2}
\noindent
One can notice that the above matrix element has an overall phase factor
$e^{ix_Lp^+(y_2^- -y_1^-)}$ as a consequence of the cancellation by the
interference processes in the left and right cut diagrams. Therefore,
the remaining contribution from the double quark scattering requires
the second quark to carry initial longitudinal momentum $x_Lp^+$ that
provides the total longitudinal momentum of the final quarks after the
elastic scattering between the leading and the second quark. The
above contributions also contain the interference between elastic
and radiative amplitudes. The leading quark in the radiative process
first emits a real gluon which converts a soft quark (which is close to real 
for small initial transverse momentum ) into the final quark with 
momentum $\ell^\prime$.

To a good approximation, especially in the case
of parton propagation in dense medium in heavy-ion collisions,
we can neglect the correlation
between the initial leading quark and the secondary quark in the
nuclear medium. The above matrix element can be factorized as a 
product of the leading quark $f_q^A(x)$ and the secondary quark
distribution in the medium. In the rest frame of the nuclear medium, we have
\begin{eqnarray}
\frac{T^{A}_{qq^\prime}(x,x_L)}{f_q^A(x)}=
\pi \int dy \rho(y)f_{q^\prime}(x_L)
\left[ 1-e^{i\frac{\ell_T^2}{2Ez(1-z)}y}\right],
\end{eqnarray}
where $y=(y_1+y_2)/2$, $f_{q^\prime}(x_L)$ is the secondary quark
distribution per nucleon and $\rho(y)$ is the spatial profile of the nucleus.

One can extend the above calculation to quark propagation in a hot
medium in high-energy heavy-ion collisions. For scattering between a 
leading quark with energy $E$ and a thermal quark with energy $\omega$,
the quark distribution in a thermal medium with quarks as the basic constituents
at temperature $T$ is,
\begin{equation}
\rho_q(y)f_q(x_L)=d_q\int \frac{d\omega}{2\pi^2} 
\frac{\omega^2}{e^{\omega/T}+1} \delta (x_L -1),
\end{equation}
where $x_L=\ell_T^2/2\omega E z(1-z)$, $d_q=4N_cN_f$ is the number of 
degrees of degeneracy for the quark
and the spatial dependence is explicitly through the temperature
$T(y)$. One can complete the integration over $z$ and $\ell_T$ in
Eq.~(\ref{eq-eloss1}). The corresponding elastic energy loss is then
\begin{eqnarray}
\Delta z_{qq}=\frac{C_F}{2N_c}\pi\frac{\alpha_s^2}{E}
d_q &&\!\!\! \int dy\int \frac{d\omega}{2\pi^2} 
\frac{\omega}{e^{\omega/T}+1}
\left[1-\cos(\omega y)\right] \nonumber \\
&\times& \left[2\ln\frac{1+\chi_\omega}{1-\chi_\omega}-
\frac{9}{4}\chi_\omega+\frac{3}{8}\chi_\omega^2\right],
\label{eq:loss2}
\end{eqnarray}
where $\chi_\omega=\sqrt{1-2\mu^2/\omega E}$ and $\mu$ 
is the transverse momentum cut-off which regularizes the collinear 
divergence normally associated with parton radiation or small angle 
scattering. In vacuum, such a cut-off separates perturbative and
non-perturbative regime of QCD. In a medium at finite temperature,
a resummation of hard thermal loops in the gluon propagator can
also provide regularization of the collinear divergence as
in previous studies \cite{Thoma:1990fm}. 

With the above equation, one can 
calculate the elastic energy loss
for any spatial profile $T(y)$ of the thermal medium. To
simplify the calculation, we neglect the thermal fluctuation
of the quark energy in the variable $\chi_{\omega}$
and replace $\omega$ with $3T$. One
can carry out the thermal averaging and obtains,
\begin{eqnarray}
\Delta z_{qq}=\frac{C_F}{2N_c}\pi\frac{\alpha_s^2}{E}
&&\!\!\! \int dy \frac{\rho_q(T)}{2T\eta(3)}
\left[\eta(2)-{\rm Re}\,\eta(2,iyT)\right] 
\nonumber \\
&\times& \left[2\ln\frac{1+\chi_T}{1-\chi_T}-
\frac{9}{4}\chi_T+\frac{3}{8}\chi_T^2\right],
\end{eqnarray}
where
\begin{eqnarray}
\eta(s,a)\equiv\sum_{n=1}^{\infty}\frac{(-1)^{n-1}}{(n+a)^s}; \;\;
\eta(s)\equiv \eta(s,0),
\end{eqnarray}
and $\rho_q(T)=d_q\eta(3)T^3/\pi^2$ is the local thermal quark density
and $\chi_T=\sqrt{1-2\mu^2/3TE}$.

One can similarly consider quark energy loss due to elastic 
quark-gluon scattering which is more involved. In particular, the
quark-gluon Compton scattering processes are very similar to the
induced gluon bremsstrahlung as considered in previous studies of
radiative energy. As we will show in a separate study \cite{wang07}, 
the original result \cite{wgdis} for radiative parton energy loss
actually contains both radiative 
and elastic processes. The LPM interference
is in fact the interference between the radiative and elastic scattering
amplitudes that are also responsible for the interference effects
in the elastic energy loss in this paper. The final result for
the elastic energy loss due to quark-gluon scattering is
\begin{eqnarray}
\Delta z_{qg}=\frac{C_A}{2N_c}\pi\frac{\alpha_s^2}{E}
&&\!\!\! \int dy \frac{\rho_g(T)}{2T\zeta(3)}
\left[\zeta(2)-{\rm Re}\,\zeta(2,iyT)\right] 
\nonumber \\
&\times& \left[3\ln\frac{1+\chi_T}{1-\chi_T}-\chi_T\right],
\end{eqnarray}
where
\begin{eqnarray}
\zeta(s,a)\equiv\sum_{n=1}^{\infty}\frac{1}{(n+a)^s}; \;\;
\zeta(s)\equiv \zeta(s,0),
\end{eqnarray}
$\rho_g(T)=d_g\zeta(3)T^3/\pi^2$ is the local thermal gluon density
and $d_g=2(N_c^2-1)$ is the number of degrees of degeneracy for gluons.
Note that $\eta(2)=\zeta(2)/2$, $\eta(3)=(3/4)\zeta(3)$, $\zeta(2)=\pi^2/6$
and $\zeta(3)\approx 1.2021$.

For the simplest parton density profile, we consider a finite medium 
with a constant parton density and a length $L$. The spatial
integrations in the above elastic energy loss can be carried out.
The total elastic energy losses are ($dE/dL=\Delta E/L$)
\end{multicols}
\begin{eqnarray}
\frac{dE_{qq}}{dL}&=&\frac{C_F}{2N_c}\frac{\pi\alpha_s^2}{24}d_q T^2
\left\{1-\frac{6}{\pi TL}\left[\frac{1}{\pi TL}-{\rm cosec}(\pi TL)
\right]\right\} \left[2\ln\frac{1+\chi_T}{1-\chi_T}-
\frac{9}{4}\chi_T+\frac{3}{8}\chi_T^2\right], \label{etot1} \\
\frac{dE_{qg}}{dL}&=&\frac{C_A}{2N_c}\frac{\pi\alpha_s^2}{12}d_g T^2
\left\{1-\frac{3}{\pi TL}\left[{\rm cth}(\pi TL)
-\frac{1}{\pi TL}\right]\right\}
\left[3\ln\frac{1+\chi_T}{1-\chi_T}-\chi_T\right], \label{etot2}
\end{eqnarray}
\begin{multicols}{2}

For a medium with infinite length $L\rightarrow \infty$, the above
elastic energy losses have finite asymptotic values,
\begin{equation}
\left(\frac{dE_{qa}}{dL}\right)_\infty=\left\langle \frac{\rho_a}{2\omega}
\int_{\mu^2}^{\omega E/2} d\ell_T^2 \frac{d\sigma_{qa}}{d\ell_T^2}
\ell_T^2 \right\rangle, (a=q,g),
\end{equation}
which can be interpreted as the average energy transfer
$\nu=\ell_T^2/2\omega$ per mean-free 
path $\lambda_{qa}=1/\rho_a\sigma_{qa}$ \cite{Wang:1996yf} due to elastic
scattering between the leading quark and thermal partons.
Here $d\sigma_{qa}/d\ell_T^2$ is the differential cross section for
$q+a$ elastic scattering.

According to Eqs.~(\ref{etot1}) and (\ref{etot2}),
\begin{eqnarray}
\left(\frac{dE_{qq}}{dL}\right)_\infty
&\approx&\frac{N_f}{6} C_F\pi\alpha_s^2 T^2 \ln\frac{3ET}{\mu^2},
 \label{etot11} \\
\left(\frac{dE_{qg}}{dL}\right)_\infty
&\approx&C_F\pi\alpha_s^2 T^2 \frac{3}{2} \ln\frac{3ET}{\mu^2}
 \label{etot22}
\end{eqnarray}
in the limit $3ET/\mu^2\gg 1$. The extra factor $3/2$ in Eq.~(\ref{etot22})
as compared to early result \cite{Thoma:1990fm} comes from the full
splitting function $[1+(1-z)^2]/z$ we used in the calculation. If we take
the approximation  $z\ll 1$ which corresponds to small angle quark-gluon
scattering in the early calculation  \cite{Thoma:1990fm}, the factor $3/2$
will disappear.

\begin{figure}
\centerline{\psfig{figure=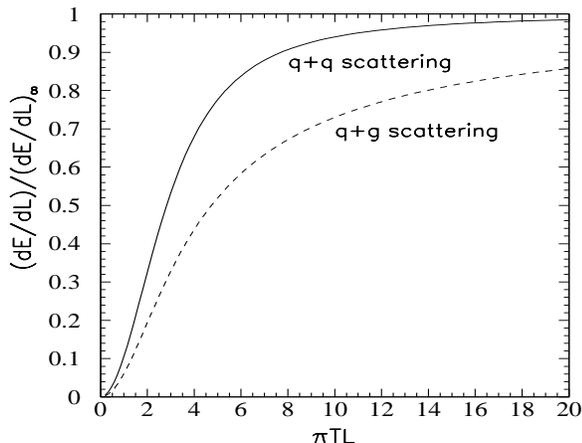,width=3.0in,height=2.3in}}
\caption{Elastic energy loss in a finite size thermal 
medium (at temperature $T$) $dE/dL$ as a function
of the medium length $L$ normalized to the asymptotic 
value $(dE/dL)_{\infty}$ for $L=\infty$.}
\label{fig2}
\end{figure}

One can observe from the above results that the effective elastic energy 
loss for a quark propagating in a medium with a finite length $L$
is reduced from its asymptotic value due to the interference between
the elastic and radiative amplitudes. The underlying physics is very
similar to the LPM interference in the gluon bremsstrahlung induced
by multiple scattering. In the elastic scattering, the gluon exchange
can be considered as the gluonic field built up around the
propagating quark over finite period of time. The formation time
in the case of elastic scattering with thermal partons is controlled by
the thermal energy of the parton and therefore is determined by the
temperature on the average. Shown in Fig.~\ref{fig2} is the quark 
elastic energy loss $dE_{qa}/dL (a=q,g)$
due to quark-quark and quark-gluon scattering normalized to the 
asymptotic value $(dE_{qa}/dL)_{\infty} (a=q,g)$ in a finite medium 
as a function of a dimensionless variable $\pi T L$. 
The dependence on $L$ is the
strongest when $\pi TL\ll 1$,
\begin{eqnarray}
\frac{dE_{qq}}{dL}\approx \frac{7}{60}(\pi T L)^2 
\left(\frac{dE_{qq}}{dL}\right)_\infty \\
\frac{dE_{qg}}{dL}\approx \frac{1}{15}(\pi T L)^2 
\left(\frac{dE_{qg}}{dL}\right)_\infty .
\end{eqnarray}
When $L\rightarrow 0$, the destructive interference becomes complete 
and the elastic energy loss vanishes.

The typical average propagation length of a parton in the most 
central $Au+Au$ collisions is about $L\sim 6$ fm. For an average
temperature $T\sim 200-300$ MeV, $\pi TL\sim 19 -28$. In this case
the interference
effect is quite small as seen in Fig.~\ref{fig2}. This is quite
different from the results in Ref. ~\cite{Peigne:2005rk}. However, for
partons produced in the peripheral region of heavy-ion
collisions whose propagation length is on the order of 1-2 fm, the
reduction of the elastic energy loss due to interference is still
significant. For an accurate and consistent treatment of elastic
energy loss, one should take into account the interference effect
and the non-trivial distance dependence. For an expanding system
with a realistic parton density profile, one has to numerically
compute the elastic energy loss and the corresponding modified
fragmentation functions according to Eq.~(\ref{eq:loss2}).
Furthermore, one should also treat radiative and elastic
energy loss in the same framework since there is intricate
connection between the two. This will be discussed in a separate 
work \cite{wang07}.

Note added: After completion of this paper, the author noticed
a recent publication \cite{Djordjevic:2006tw} in which the finite size 
effect on elastic energy loss is a result of the imposed finite interaction
time which modifies the energy conservation and therefore the phase space
available for final partons after elastic scattering. The interference
effect discussed in the current paper is not included and therefore 
the obtained length and energy dependence of the elastic energy loss 
are different.

The author would like to acknowledge useful discussions with M. Djordjevic and
M. Gyulassy during the write-up of this manuscript.
He also thanks S. Peigne for reading and comments on the manuscript. 
This work is supported  by the Director, Office of Energy
Research, Office of High Energy and Nuclear Physics, Divisions of 
Nuclear Physics, of the U.S. Department of Energy under Contract No.
DE-AC02-05CH11231.


\end{multicols}

\end{document}